\documentclass[11pt]{article} 
\usepackage{hyperref} 
\pdfoutput=1 
\begin{document} 
\title{The Brine Shrimp's Butterfly Stroke} 
\author{Brennan Johnson\textsuperscript{1}, Deborah Garrity\textsuperscript{2}, and Lakshmi Prasad Dasi\textsuperscript{1,3} \\ 
\\\vspace{6pt} 1. School of Biomedical Engineering, \\ 2. Department of Biology, \\3. Department of Mechanical Engineering, \\\\ Colorado State University, Fort Collins, CO 80523, USA} 
\maketitle 

\section{Introduction} 
We investigate the fluid dynamics of brine shrimp (Artemia) in the nauplius stage of swimming. These larvae swim using two arms that oscillate in a “butter fly stroke” manner at 6-10Hz. Previous studies have been performed on larval brine shrimp swimming\textsuperscript{1-3}. These studies indicated that the propulsion dynamics was largely dominated by viscous forces with a mechanism described as “rowing”. A more recent study attempted to measure the vorticity dynamics and captured a single vortex ring system\textsuperscript{3}. However, these earlier studies lacked the spatial and temporal resolution to resolve the smallest scales of motion along with any dynamics effects. In this fluid dynamics video study, we use a time-resolved particle image velocimetry (PIV) technique with nano-particles as seeding material to examine the fluid dynamics at a much higher spatio-temporal resolution compared to earlier studies, with particular emphasis on resolving the vorticity dynamics of the propulsive mechanisms utilized by the larvae. 

\section{Results} 
Two- to three-day-old brine shrimp were filmed while freely swimming in a solution of water and hydroxyapatite nano-powder, used as seeding material. The images were captured with a CMOS camera at a sampling rate of 1KHz. Image pre-conditioning was conducted to remove background artifacts,  followed by cross-correlation interrogation analysis to obtain particle image velocimetry measurements of the 2D velocity field. The brine shrimp were viewed from above in order to capture the plane that lies along the body and arms. The spatial resolution of the resulting regularly spaced 2D vector field grid size was approximately 50 $\mu$m.  Based on the peak forward velocity of the larvae, the body length, and the oscillation frequency, the Reynolds and Womersley numbers were found to be 8 and 5 respectively. 

Vorticity dynamics reveals the formation of a vortex ring structure at the tip of each arm at the beginning of the power stroke. This two vortex ring structure evolves dramatically with time as the stroke progresses. Most interestingly, the outer circulation weakens while the inner circulation strengthens over the power stroke. In addition, the outer vortices translate backwards while stretching from the induced fields of the inner vortices. Ultimately the outer vortices seem to be entrained into the inner strong vortices resembling a leap-frog phenomena of a two-vortex ring system. The gaining strength of the inner vortex correlates well with the acceleration and forward movement of the larvae. We also investigated the motion of a one-arm brine-shrimp larvae which also showed the dynamics of only a single vortex system. The resultant motion was therefore a circular trajectory for the larvae.

\section{References} 
\begin{enumerate} 
\item 	Williams TA. A model of rowing propulsion and the ontogeny of locomotion in artemia larvae. Biological Bulletin. 1994; 187:164-173.
\item 	Williams TA. Locomotion in developing artemia larvae - mechanical analysis of antennal propulsors based on large-scale physical models. Biological Bulletin. 1994; 187:156-163.
\item 	Mochizuki O, Asme. Micro vortex flow induced by small life; 2007.
\end{enumerate}

\end{document}